\documentstyle[12pt]{article}
\newcommand{\bq}{\begin{equation}}
\newcommand{\ee}{\end{equation}}

\newcommand{\fr}[2]{\frac{#1}{#2}}

\begin{document}
\pagestyle{plain}
\pagenumbering{arabic}

\begin{flushright}
{\large Preprint BUDKERINP 95-86}\\
hep-ph/9510343\\
October-November 1995\\
\end{flushright}

\begin{center}{\Large \bf The Status of Renormalon }\\

\vspace{0.5cm}

{\bf S. V. Faleev,}\footnote{e-mail address:
S.V.Faleev@INP.NSK.SU}
{\bf  P. G. Silvestrov}\footnote{e-mail address:
P.G.Silvestrov@INP.NSK.SU}
 \\ Budker Institute of Nuclear
Physics, 630090 Novosibirsk, Russia

\vspace{0.5cm}

\end{center}
\begin{abstract}

It is shown that the series of renormalon--type graphs, which
consist in the chain of insertions to one soft(hard)
gluon(photon) line is in fact ill defined. Each new type of
insertions, which appears in the higher orders of perturbation
theory, generates the correction to renormalon of the order of
$\sim 1$. However, this series of the corrections to the
asymptotics although have no small parameter but hopefully is
not the asymptotic one. The consideration based on the use of
the renormalization group equation for effective charge is
supported by the direct diagrammatic picture.

\end{abstract}

\newpage

{\bf 1.} For many years it was generally accepted that the true
asymptotics of the perturbation theory in theories with running
coupling constant is determined by the renormalons
\cite{t'Hooft,ren}. Nowadays, the renewed interest is
demonstrated in asymptotic estimates of perturbative series
[3-11].
%\cite{West,Brow,VZre,Grun,Broa,Maxwel}.
It results even in the recent
attempts \cite{BBB} to use the renormalon for direct calculation
of experimentally measurable quantities.

However, the accurate determination of renormalon-type
asymptotics appears to be not so simple problem. It was
recognized by the specialists \cite{Grun,Mueller,Beneke} that
the overall normalization factor of the renormalon could not be
found without taking into account of all terms of the expansion
of, say, the Gell-Mann--Low function. However, the usual proof
of this fact do not refer on the direct counting of the Feynman
graphs (see also the discussion after our equation (\ref{b2nn})).
Also recently Vainshtein and Zakharov \cite{Vainshtain} have
found the new source of uncertainty, which makes difficult the
quantitative finding of
the renormalon asymptotics (at least for the ultraviolet
renormalon). The generally considered renormalon chain of graphs
is formed by dressing of one gluon(photon) line by
various insertions. The authors of ref. \cite{Vainshtain} have
broken this tradition and considered the contribution to the
ultraviolet renormalon of diagrams with two, three etc. dressed
lines. Their analysis based on the operator product expansion,
showed that contributions to the asymptotics from diagrams with
any few dressed gluon lines are all of the same order of
magnitude and also are parametrically larger than the usual
renormalon with only one dressed line.
%Thus, there is no clear
%way now to find the overall amplitude of the ultraviolet
%renormalon.

In the present paper we would like to continue the
analysis of the renormalon--type chain of Feynman diagrams.
We consider the dressing of single gluon(photon) line as
for the traditional renormalon, but try to estimate the role of
the arbitrary high order insertions to this "dressed" gluon. It
will be shown that each new type of insertions generates the
correction to renormalon of the order of $\sim 1$.  Thus
in order to find the overall normalization
of the asymptotics one has to calculate all the coefficients
$b_0,b_1,b_2, \ ...$ of the expansion of the renormalization
group equation for, say, effective charge (see the eq.
(\ref{GML}) below).  Our approach is equally valid for both
infrared and ultraviolet renormalons. However, because there
exists another source of problems for the ultraviolet
renormalon \cite{Vainshtain}, we will
concentrate our attention on the infrared one.

It is a tradition now to consider the renormalon for QED. In
this paper we will also discuss only the QED--type diagrams of
the perturbation theory, without the self-interaction of gluons.
However, while considering the renormalization group
coefficients we would like to use the QCD values $b_0 =
\fr{1}{4\pi} \left( \fr{11}{3}N_c - \fr{2}{3} N_f\right)$, $b_1
=\left( \fr{1}{4\pi}\right)^2 \left( \fr{34}{3}N_c^2 -
\fr{13}{3} N_f N_c + N_f/N_c \right), \ ... $ . Thus by hand we
avoid the problem of the gauge--independent definition of the
renormalon chain in QCD.

{\bf 2.} The contribution of the diagrams with exchange of one
soft gluon(photon) to some "physical" quantity (say $R_{e^+ e^-
\rightarrow hadrons}$ , or the correlator of two currents) has
the generic form
\bq\label{ren}
R = \int_{k\ll Q} \alpha(k) \fr{k^2 dk^2}{Q^4} \ \ .
\ee
This is the first infrared renormalon, because of the
contribution proportional to $dk^2/Q^2$ vanishes due to the
gauge invariance.  The Feynman graphs corresponding to this
value are shown in fig.~1. More precisely the graphs of fig.~1
correspond to (\ref{ren}) with the running coupling constant
$\alpha(k)$ replaced by the fixed value $\alpha_0=\alpha(Q)$~.
In (\ref{ren}) we have written down the effective running
coupling constant $\alpha(k) = \alpha_{eff}(k)$ instead of
$\alpha_0$. $\alpha_{eff}(k)$ is trivially connected with the
transverse part of the gluon propagator $G_{\mu\nu}(k)$.  The
function $\alpha(k)$ satisfies the renormalization group
equation:
\begin{eqnarray}\label{GML}
\fr{d\alpha}{dx}
 &=&
b_0 \alpha^2 +b_1 \alpha^3 +b_2 \alpha^4 + ... \ \ , \\
x&=& \ln\left( \fr{Q^2}{k^2} \right) \ \ \ , \nonumber
\end{eqnarray}
where for practical applications we will use $N_f=N_c=3 \ , \
4\pi b_0=9 \ , \ 16\pi^2 b_1 =64$ . Starting from $b_2$ the
coefficients of the expansion (\ref{GML}) depend on the
renormalization scheme. It is to be noted here that we have
fixed the renormalization scheme by considering the effective
charge. Thus our coefficients $b_2,b_3,...$ are neither the free
parameters, nor the known, say, for $\overline{MS}$ scheme,
$b_2(\overline{MS}), \ b_3(\overline{MS})$ .

In order to find the leading contribution to the asymptotics,
one may neglect $b_1,b_2,$ etc. in (\ref{GML}). Now instead of
(\ref{ren})
\begin{eqnarray}\label{rendef}
R &=& \fr{2}{\alpha_0} \int_0^{\infty} \alpha(x) e^{-2x}dx =
\int_0^{\infty}
\fr{e^{-2x}}{1-b_0\alpha_0 x} 2 dx = \nonumber \\
&=& \int_0^{\infty} \sum_{N=0} ( b_0\alpha_0 x)^N
e^{-2x} 2 dx = \sum_{N=0} \left(
\fr{b_0\alpha_0}{2} \right)^N N! \ \ \ .
\end{eqnarray}
Here the first equality is also the explicit definition of the
quantity, which we are going to consider in this paper.

It is seen immediately from (\ref{rendef}) that if $b_0 > 0$,
our renormalon is ill defined due to the Landau pole. However,
in this paper we will consider only the asymptotics of the
perturbation theory and will not concern the issue of the
nonperturbatve definition of the integral (\ref{rendef}).

The way of calculation of the renormalon now seems
straightforward. One should find, step by step, the function
$\alpha(x)$ from (\ref{GML}), substitute it into (\ref{rendef}),
and look for the asymptotics. Nevertheless, before passing to
this formal way, let us illustrate the role of complicated
contributions to the renormalon by the explicit estimate of
corresponding Feynman graphs.

The fig.~2 shows the chain of diagrams corresponding to
renormalization of the soft gluon line in the leading
approximation (\ref{rendef}).  As we have said above, we show
only the QED--type diagrams consisting of fermionic bubbles
without gluon self-interaction. In the $N$-th order of
perturbation theory each of the $N$ bubbles from fig.~2
generates the factor
\bq\label{b0}
b_0\alpha_0 \ln\left( \fr{Q^2}{k^2} \right)
\ee
in the integrand of (\ref{ren}),(\ref{rendef}). Now the
difference between theories (QCD,QED) is hidden in the factor
$b_0$, accompanying the single bubble.

Now let us replace two of the simple bubbles by the more
complicated two loop diagram, as it is shown in fig.~3.  Again
we have shown only the QED diagram. Moreover we have shown
explicitly only one of the two diagrams (compare with fig.~1) of
the second order in $\alpha_0$. As it is indicated on the
figure, the -- two loop bubble generates the factor
\bq\label{b1}
b_1 \alpha_0^2 \ln\left( \fr{Q^2}{k^2} \right)
\ee
in the integrand, which has one power of large logarithm less
(or one $\alpha_0$ more) than the leading order contribution
(\ref{rendef}). On the other hand a large combinatorical factor
$N-1$ appears due to a number of permutations of the second
order bubble among the simple bubbles. As a result one has
\bq\label{b1sum}
(N-1) b_1 \alpha_0^2 \ln\left( \fr{Q^2}{k^2} \right) \left[ b_0
\alpha_0 \ln\left( \fr{Q^2}{k^2} \right) \right]^{N-2}
\rightarrow \left( \fr{b_0\alpha_0}{2}\right)^N N!
\fr{2b_1}{b_0^2} \ .
\ee
Thus we can see that taking into account one second order
insertion into the soft gluon line leads to the correction of
the order of one to the trivial asymptotics (\ref{rendef}).
Summation over the number of the second order insertions shown
in fig.~3 leads to a simple exponentiation of this correction
\bq\label{b1exp}
\left( \fr{b_0\alpha_0}{2}\right)^N N!
\exp \left( \fr{2b_1}{b_0^2} \right) \ .
\ee

In last few years it became very popular \cite{Broa,Maxwel} to
consider the renormalon in the limit $N_f \rightarrow \infty$.
{}From this point of view the correction (\ref{b1sum}) is nothing
more than the $\sim 1/N_f$ correction, because of $2b_1/b_0^2
\sim 1/N_f \ll 1$ . However for practically interesting $N_c=3$
and $N_f=3,4,5$ one has, respectively, ${2b_1}/{b_0^2}
=\fr{128}{81}\, , \, \fr{906}{625} \, , \, \fr{678}{529} $ .

Consider now the more complicated diagram of fig.~4 with
dressing of the internal gluon line of the second order bubble.
At this point it is natural to write down explicitly the last
integration over internal momentum of the two loop diagram
\bq\label{b1l}
b_1 \alpha_0^2 \ln\left( \fr{Q^2}{k^2} \right) = b_1 \alpha_0^2
\int_{k^2}^{Q^2} \fr{dq^2}{q^2} \ .
\ee
Now the dressing of the gluon line evidently leads to
\bq\label{b1ll}
b_1 \alpha_0^2 \int_{k^2}^{Q^2} \left[ b_0
\alpha_0 \ln\left( \fr{Q^2}{q^2} \right) \right]^n
\fr{dq^2}{q^2} = \fr{1}{n+1} b_1 \alpha_0^2 \ln\left(
\fr{Q^2}{k^2} \right) \left[ b_0
\alpha_0 \ln\left( \fr{Q^2}{k^2} \right) \right]^n \ .
\ee
Thus up to the overall factor $\fr{1}{n+1}$ the contribution of
diagram of fig.~4 coincides with that of fig.~3. Summation over
the number of simple bubbles inserted into the large bubble --
$n$ naturally leads to $\ln(N)$. Also taking into account
a number of large bubbles of fig.~4 allows to exponentiate the
correction
\bq\label{b1lexp}
\left( \fr{b_0\alpha_0}{2}\right)^N N!
\exp \left( \fr{2b_1}{b_0^2} \ln(N) \right) = \left(
\fr{b_0\alpha_0}{2}\right)^N N^{\fr{2b_1}{b_0^2}} N! \ .
\ee
This is the generally recognized expression for the infrared
renormalon. Our arguments up to this stage simply repeat the
line of reasoning of the paper \cite{VZre}, though may be in
more details.

It is clear, that the argument of the exponent in (\ref{b1lexp})
was found with the $\sim~1/\ln(N)$ accuracy and therefore the
nontrivial overall factor as well as the function of $N$, weaker
than $N^{\gamma}$, may appear in (\ref{b1lexp}), as we consider
in detail in the following section.

Now let us consider the three loop correction to the renormalon,
shown in fig.~5.
Like it was done before, we show only one example of
the third order diagram. All these contributions generate the
factor in the integrand of (\ref{rendef})
\bq\label{b2}
b_2 \alpha_0^3 \ln\left( \fr{Q^2}{k^2} \right) \ .
\ee
Thus here we have two extra $\alpha_0$, which at first glance
could not be compensated by one combinatorical factor $N$ and
hence the diagram of fig.~5 seems to generate only the $\sim
1/N$ correction to renormalon. In particular such conclusion
was drown by Zakharov \cite{VZre}. However, let us see,
what happens
if one dresses the internal gluon lines of the three loop
diagram (fig.~6). Now summation over the number of trivial
corrections inserted into a large diagram $n_1,n_2$ gives:
\begin{eqnarray}\label{b2nn}
&\,& b_2 \alpha_0 \ln\left( \fr{Q^2}{k^2} \right)
\sum_{n_1,n_2}
\alpha_0^2 \fr{1}{n_1+n_2+1} (N-n_1-n_2-2) \nonumber \\
&\sim& b_2 \alpha_0 \ln\left( \fr{Q^2}{k^2} \right) \times
(\alpha_0 N)^2 \sim b_2 \alpha_0 \ln\left( \fr{Q^2}{k^2} \right)
\ .
\end{eqnarray}
Here the factor $(n_1+n_2+1)^{-1}$ in the sum appears after
integration over the internal momentum of the large bubble (see
(\ref{b1l})), while the factor $(N-n_1-n_2-2)$ accounts for the
number of permutations of our large bubble with simple small
bubbles on the main gluon line.
So we see that after dressing of all gluon
lines the three loop ($\sim b_2$) diagram generates the
correction to renormalon of the order of $\sim 1$. One can
easily show that four loop ($\sim b_3$), five loop ($\sim b_4$)
etc. diagrams generate the corrections of the same order of
magnitude. Previously the analogous proof of the importance
of the high loop corrections to the renormalon chain was done
by Mueller \cite{private}, but this result was not published.
%In general, this conclusion is the main result of our
%paper and in the following we will only rederive it in more
%formal way.

{\bf 3.} In this section we would like to turn back to the
direct analysis of the renormalon (\ref{ren},\ref{rendef})
without referring to the Feynman diagrams. It is easy to
integrate formally the renormalization group equation
(\ref{GML})
\bq\label{GMLi}
-\fr{1}{\alpha}+\fr{1}{\alpha_0}-
\fr{b_1}{b_0}\ln\left( \fr{\alpha}{\alpha_0} \right) -
c_2 (\alpha-\alpha_0)-c_3 \fr{\alpha^2-\alpha_0^2}{2}-...
=b_0 x \ ,
\ee
where, $c_2=b_2/b_0-b_1^2/b_0^2 \ , \
c_3=b_3/b_0-2b_2 b_1/b_0^2+b_1^3/b_0^3 \ ...$
. Also it is convenient to introduce the new set of variables
\begin{eqnarray}\label{set}
t&=& 2x \ , \nonumber \\
\fr{b_0\alpha}{2} &\rightarrow& \alpha \ , \
\fr{b_0\alpha_0}{2} = a  \\
\beta_1 &=& \beta = \fr{2b_1}{b_0^2}  \ , \ \beta_n = \left(
\fr{2}{b_0}\right)^n c_n
 \ ... \ . \nonumber
\end{eqnarray}
Now the renormalon (\ref{rendef}) takes the form
\begin{eqnarray}\label{renalpha}
R &=& \int_0^{\infty} \fr{\alpha}{a} e^{-t} dt \ \ , \\
\fr{\alpha}{a} &=& \fr{1}{1-at- \beta_1 a\ln (
{\displaystyle {\alpha}/{a}} ) - \beta_2(\alpha-a)a
-\beta_3(\alpha^2-a^2)a - ...} \ \ . \nonumber
\end{eqnarray}

Although we have shown, that the corrections to renormalon
generated by the high order contributions to the renormalization
group equation $b_2\alpha^4,b_3\alpha^5...$ (\ref{GML}) are not
small, the correction induced by the second term $b_1\alpha^3$
still plays an outstanding role due to the additional
enhancement by $\ln(N)$ (see (\ref{b1lexp})). Therefore one
should be more careful, while treating these enhanced
contributions. Moreover, the iterations of the second order
diagram of fig.~3, e.g. as it is shown in fig.~7, may also have
some additional enhancement. Thus at the first stage let us omit
$\beta_2, \beta_3, \beta_4 \ldots$ in (\ref{renalpha}) and
consider the truncated effective charge:
\bq\label{renalb1}
\fr{\alpha}{a} = \fr{1}{1-at- \beta a\ln (
{\displaystyle {\alpha}/{a}} ) } \ \ .
\ee
This is the transcendental equation for the function
$\alpha=\alpha(a)$, which may be solved iteratively. For
estimate of the $N$-th order of perturbation theory we will
often use the formula for $N$-th term of the expansion of the
integral in powers of $a$
\bq\label{formula}
\left\{ \int \fr{e^{-t}dt}{1-at} \left[ \fr{\beta a}{1-at}
\right]^k \left[ \ln \fr{1}{1-at}  \right]^m
\right\}_{N} = a^N N! \fr{\beta^k}{k!} \left[ \ln
\fr{N}{k} \right]^m \left( 1+ O\bigg( \fr{1}{\ln N}
\bigg) \right) \ .
\ee
Here both $m$ and $k$ are supposed to be large $m,k\sim \ln (N)$
. Everywhere in this section we suppose that not only the $N$,
but also the $\ln (N)$ is a large parameter and neglect all the
corrections of the order of $\sim 1/\ln(N)$. In order to derive
the formula (\ref{formula}) one has to use the asymptotics
of gamma-function as well as the trivial identity
\bq\label{vareps}
\big( \ln(p) \big)^n = \lim_{\varepsilon \rightarrow 0} \left(
\fr{\partial}{\partial\varepsilon} \right)^n p^{\varepsilon} \ .
\ee
Now it is easy to make the first iteration in
(\ref{renalb1},\ref{renalpha})
\begin{eqnarray}\label{R1}
\{ R_1\}_N = \int \fr{e^{-t}dt}{1-at+ \beta a\ln (
{\displaystyle {1-at} ) }} &=& \int
\fr{e^{-t}dt}{1-at} \sum_p \left[ \fr{\beta a}{1-at}
\ln \fr{1}{1-at}  \right]^p = \\
= a^N N! \sum_p \fr{1}{p!} \left[ \beta \ln \bigg( \fr{N}{p}
\bigg) \right]^p &=& \left( \fr{b_0 \alpha_0}{2} \right)^N
\left[
\fr{N}{\beta\ln N} \right]^{\beta} N! \left( 1+ O\bigg(
\fr{1}{\ln N} \bigg) \right) \ . \nonumber
\end{eqnarray}
Here for brevity we "forget" to indicate, that only the
$N$-th term of the expansion of both integrals over $a$ is
considered. The asymptotics (\ref{R1}) should be compared with
(\ref{b1lexp}). One can see that the consistent treatment of
the diagrams of fig.~4, which in fact we have done in
(\ref{R1}), results in the nontrivial small factor $(\ln
N)^{-\beta}$ as compared to the naive result (\ref{b1lexp}).

Consider now the effect of iteration of two loop correction
(fig.~7), which in terms of $\alpha$ means
\bq\label{iter}
\left( \fr{a}{\alpha} \right)_{second \ iteration} = 1-at+\beta
a \ln (1-at +\beta
a \ln (1-at)) \ .
\ee
To this end one has simply to replace in (\ref{R1})
\begin{eqnarray}\label{replace}
&\,&\left[ \fr{\beta a}{1-at}
\ln \fr{1}{1-at}  \right]^p \rightarrow
\left[ \fr{-\beta a \ln (1-at +\beta
\ln (1-at))}{1-at} \right]^p = \nonumber \\
&=& \left[ \fr{\beta a}{1-at} L \right]^p
\left\{ 1- \fr{1}{L} \ln \left( 1- \fr{\beta a}{1-at} L\right)
\right\}^p \ ,
\end{eqnarray}
where we introduce $L= \ln\left( \fr{1}{1-at} \right)$. As
we have seen from (\ref{formula},\ref{R1}), from the point of
view of perturbation theory $\fr{\beta a}{1-at} \approx (\ln
N)^{-1}$, while $L \approx \ln N$. As a result the expression in
curly brackets in (\ref{replace}) is of the form
\bq\label{esmall}
1- \fr{1}{L} \ln \left( 1- \fr{\beta a}{1-at} L\right)
\sim 1 - O\left( \fr{\ln(\ln N))}{\ln N} \right)
\ee
and may be written as the exponent of a small quantity. Thus
instead of (\ref{replace}) one gets
\bq\label{replace1}
\left[ \fr{\beta a}{1-at} L \right]^p
\left[1 - \fr{\beta a}{1-at} L \right]^{-\fr{p}{L}} =
\sum_{n=0} \left[ \fr{\beta a}{1-at} L \right]^{p+n}
\fr{n^{\fr{p}{L}-1}}{\Gamma(p/L)} \ .
\ee
In terms of Feynman graphs the variables $p$ and $n$ are
respectively the number of large second order bubbles and
number of internal second order bubbles (see fig.~7). Also all
wavy lines in fig.~7 are supposed to be renormalized by
arbitrary number of simple one loop insertions (like in fig.~2).

Substitution of (\ref{replace1}) into (\ref{R1}) gives
\bq\label{R2}
\{ R_2 \}_N =a^N N! \sum_{q=0}^{\infty} \sum_{n=0}^q \fr{1}{q!}
\left[ \beta \ln \left( \fr{N}{q} \right) \right]^q
\fr{n^{\fr{q-n}{\ln N}-1}}{\Gamma \left( \fr{q-n}{\ln N}
\right)}
\ .
\ee
Here we have denoted $p+n$ (\ref{replace1}) by $q$. Due to
$[\beta \ln N]^q/q!$ the series in $q$ (\ref{R2}) has a narrow
peak. Therefore, up to $\sim 1/\ln{N}$ corrections
\bq\label{R2i}
\{ R_2 \}_N =\left( \fr{b_0 \alpha_0}{2} \right)^N
\left[ \fr{N}{\beta} \right]^{\beta} N! \int_0^{\beta}
x^{\beta-x-1} e^{-x\ln(\ln N)} \fr{dx}{\Gamma(\beta-x)} \ ,
\ee
where $x=n/\ln (N)$ (\ref{R2}).
The explicit integration here may be performed only if the
$\ln(\ln N) \gg 1$
\bq\label{R2e}
\{ R_2 \}_N \approx \left( \fr{b_0 \alpha_0}{2} \right)^N
\left[
\fr{N}{\beta\ln(\ln N)} \right]^{\beta} N! \left( 1+ O\bigg(
\fr{1}{\ln(\ln N)} \bigg) \right) \ .
\ee
Thus again we can see (compare (\ref{R2e}) and (\ref{R1})) that
taking into account the diagrams of fig.~7, or in other words,
making the second iteration (\ref{iter}) in the transcendental
equation (\ref{renalb1}), leads to parametrically large
renormalization of the renormalon.

In the analogous way one may perform the third iteration of
(\ref{renalb1}). The result reads
\bq\label{R3i}
\{ R_3 \}_N =\left( \fr{b_0 \alpha_0}{2} \right)^N
\left[ \fr{N}{\beta} \right]^{\beta} N! \int_0^{\beta} dx_1
\int_0^{x_1} dx_2
\fr{(x_1-x_2)^{\beta-x_1-1}}{\Gamma(\beta-x_1)}
\fr{x_2^{x_1-x_2-1}}{\Gamma(x_1-x_2)} e^{-x_2\ln(\ln N)}  \ .
\ee
Here the integration may be performed analytically only if the
$\ln (\ln (\ln N)) \gg 1$ resulting in the same expression as
(\ref{R2e}), but with $\ln (\ln N)$ replaced by $\ln (\ln (\ln
N))$ both in the main formula and in the estimate of the error.

For the case of arbitrary $k+1$ iterations in (\ref{renalb1})
the generalization of (\ref{R2i}) and
(\ref{R3i}) leads to
\begin{eqnarray}\label{Rk}
&\,& \{ R_{k+1} \}_N =\left( \fr{b_0 \alpha_0}{2} \right)^N
\left[ \fr{N}{\beta} \right]^{\beta} N! \int_0^{\beta} dx_1
\int_0^{x_1} dx_2 \, ... \, \int_0^{x_{k-1}} dx_k
\fr{\beta-x_1}{\Gamma(1+\beta-x_1)} \\ &\,& \times
\fr{(x_1-x_2)^{\beta-x_1}}{\Gamma(1+x_1-x_2)}
\fr{(x_2-x_3)^{x_1-x_2}}{\Gamma(1+x_2-x_3)} \, ... \,
\fr{(x_{k-1}-x_k)^{x_{k-2}-x_{k-1}}}{\Gamma(1+x_{k-1}-x_{k})}
\fr{x_k^{x_{k-1}-x_k}}{x_k} e^{-x_k\ln(\ln N)}  \ . \nonumber
\end{eqnarray}
Again one can calculate this integral analytically if the $k$-th
logarithm is large \\ $\ln (\ln (\, ... \, \ln N)\, ... \,) \gg
1$, though the expression (\ref{Rk}) itself has the accuracy
$\sim 1/\ln N$.

Looking at the formula (\ref{Rk}), one may even doubt, whether
it has any finite limit for large $k$. In order to prove this
finiteness consider in more details the last integral in
(\ref{Rk})
\begin{eqnarray}\label{kint}
&\,&\int_0^{x_{k-1}} dx_k \fr{(x_{k-1}-x_k)^{x_{k-2}-x_{k-1}}
}{\Gamma(1+
x_{k-1}-x_k)} \fr{x_k^{x_{k-1}-x_k}}{x_k}  e^{-x_k\ln(\ln N)} =
\\ &\,&
x_{k-1}^{x_{k-2}-x_{k-1}-1} \bigg( 1+
\sum_{m+n>0} c_{mnl} x_{k-2}^m x_{k-1}^n (\ln x_{k-1})^l \bigg)
\ . \nonumber
\end{eqnarray}
Substitution of this result into (\ref{Rk}) naturally leads to
the following estimate
\bq\label{Rkk}
\{ R_{k+1} \}_N = \{ R_{k} \}_N
\left( 1+ O\bigg(
\fr{const^k}{k!} \bigg) \right) \ .
\ee
By the way we have shown here that the integral in (\ref{Rk})
at large $k$ forgets effectively about the value of $\ln(N)$~.

Thus we arrive at the surprising result. If one performs only
one (a few) iterations while solving the transcendental equation
for the effective charge (\ref{renalb1}), the asymptotics of the
perturbation theory differs drastically
(\ref{R1},\ref{R2i}--\ref{R3i}) from the generally
recognized renormalon (\ref{b1lexp}). Nevertheless, after taking
into account the infinite number of iterations (\ref{Rk}) the
renormalon (\ref{b1lexp}) is restored up to some overall
constant.

{\bf 4.} In the present paper we have no plan to perform the
complete analysis of the contribution to renormalon from the
high order terms of the renormalization group equation
(\ref{GML}) $b_2\alpha^4, b_3\alpha^5 \ ...$ (in
particular because the
coefficients $b_2,b_3$ themselves are not known). It will be
enough for us to show that all the corrections to renormalon
due to these high order terms are equally important. To this end
we will consider the contribution to renormalon from the
multi-loop corrections in the equation (\ref{GML}) only in the
leading nontrivial approximation.

As we have seen (\ref{GMLi}--\ref{renalpha}) in the formal
solution of the renormalization group equation the coefficients
$b_n$ have been converted into $\beta_n = 2^n b_n/b_0^{n+1} + \
...$ . In the linear in $\beta_2, \beta_3, ....$ approximation
the effective charge (\ref{renalpha}) takes the form
\bq\label{alphal}
\fr{\alpha}{a} = \fr{\alpha_1}{a} \bigg( 1+\beta_2\alpha_1^2 +
\beta_3\alpha_1^3 + \, ...\, \bigg) \ ,
\ee
where $\alpha_1$ is the solution of the equation
(\ref{renalb1}). While going from (\ref{renalpha}) to
(\ref{alphal}) we have ignored the $\beta_2 a^2$ in comparison
with $\beta_2 \alpha^2$, $\beta_3 a^3$ in comparison with
$\beta_3 \alpha^3$ and so on. With the experience of the
previous section one can easily show, that the omitted
contributions are of the order of $\sim 1/N$.

Now it is easy to repeat all the logics, which led us to the
equation (\ref{Rk})
\begin{eqnarray}\label{Rkb}
&\,& \{ R_{k+1} \}_N =\left( \fr{b_0 \alpha_0}{2} \right)^N
\left[ \fr{N}{\beta} \right]^{\beta} N! \int_0^{\beta} dx_1 \\
&\,& \left( 1+ \fr{\beta_2}{2!}(\beta -x_1)^2 +
\fr{\beta_3}{3!}(\beta -x_1)^3 + \, ... \, \right)
\int_0^{x_1} dx_2 \, ... \ \ , \nonumber
\end{eqnarray}
where the part of the formula after integration over $dx_2$
simply repeats the corresponding part of the equation
(\ref{Rk}). Keeping in mind the finiteness of both (\ref{Rk})
and (\ref{Rkb}) (see (\ref{kint}) and (\ref{Rkk})) one can see
that all $\beta_2, \beta_3, \, ...$ make contributions of the
order of $\sim 1$ to the renormalon.

Nevertheless the formula (\ref{Rkb}) allows one to draw not
only the pessimistic conclusions. All the high order ($\sim
\beta_n$) terms in (\ref{Rkb}) have appeared in the combination
$\beta_n/n!$~. Thus one may hope, that at least the series of
the corrections to the renormalon is not the asymptotic one.

{\bf 5.} The formulas like (\ref{Rk}) still are too complicated
for practical use. In this section we would like to develop
another method for calculation of the renormalon--type
asymptotics, which at least enables to perform the simple
numerical computations.

Let us again introduce the new variables instead of (\ref{set})
\bq\label{set2}
f=\fr{\alpha}{a}  \ , \ s=at \ , \
\gamma= \fr{2b_1}{b_0^2 t} = \fr{2b_1}{b_0^2 N} \ .
\ee
Here in the last equality we use, that in all corrections to
the trivial renormalon chain (\ref{rendef}) the variable $t$ may
be replaced by the saddle--point value $t=N$ up to corrections
of the order of $\sim [\ln N]^2/N$. Now the truncated formula
for the effective charge (\ref{renalb1}) takes the form
\bq\label{alphanew}
f = \fr{1}{1-s-\gamma s\ln{f}} \ ,
\ee
where $\gamma \sim 1/N$ . The $N$-th term of the expansion of
$f$ in the series over $s$ just gives the asymptotics
\begin{eqnarray}\label{serasym}
f &=& 1 + \sum_1^{\infty} C_n(\gamma) s^n \ , \\
\{ R \}_N &=& \left( \fr{b_0 \alpha_0}{2} \right)^N
 N! C_N(\beta/N) \ \ . \nonumber
\end{eqnarray}
It is easy to show that $f(s)$ satisfies the differential
equation
\bq\label{difeq}
f'=\fr{df}{ds} = \fr{f(f-1)}{s} +\gamma s
f f' \ ,
\ee
which leads to the recursion relation for the coefficients of
the expansion (\ref{serasym})
\bq\label{recur}
C_{k+1} = \fr{1}{k} \sum_{n=1}^k C_n C_{k+1-n} +\gamma C_k
+\fr{\gamma}{2} \sum_{n=1}^{k-1} C_n C_{k-n}  \ ,
\ee
with the initial condition $C_1=1$. For analytical
investigation this formula is even more complicated, than the
original equations (\ref{renalpha}), (\ref{renalb1}). On the
other hand, it is easy to solve (\ref{recur}) numerically.

We have performed the numerical simulations with $b_0=9, \
b_1=64$ and $N$ up to $\sim 16000$. The result reads
\bq\label{numerics}
R=0.27 \sum \left( \fr{9}{2} \alpha_0 \right)^N N^{\fr{128}{81}}
N! \ \ .
\ee
Let us recall that this result is obtained for
$\beta_2=\beta_3=..=0$ (see (\ref{set}),(\ref{renalpha})) or, in
other words, we were interested in the solution of the equation
(compare with (\ref{GML}))
\bq\label{betaf}
\fr{d\alpha}{dx} = \fr{b_0 \alpha^2}{1 - {\displaystyle
\fr{b_1}{b_0} \alpha}} \ .
\ee
If one expands the r.h.s. of this equation in series in
$\alpha$, the terms proportional to $b_1^2, b_1^3, \ ...$ will
mimic effectively the $\sim b_2, b_3, \ ...$ terms of the full
equation (\ref{GML}). On the other hand, one may consider the
pure truncated renormalization group equation:
\bq\label{betaf1}
\fr{d\alpha}{dx} = b_0 \alpha^2 + b_1 \alpha^3 \ .
\ee
In this case the formulas (\ref{alphanew}), (\ref{difeq}) are
modified
\bq\label{alphanewnew}
f = \fr{1}{1-s-\gamma s\displaystyle{ \ln\bigg(
\fr{f}{1 + \gamma sf}\bigg) }} \ , \ \ \
f'= \fr{f(f-1)}{s} +\gamma sf^2 (f-1-\gamma s f) \ .
\ee
More precisely this formula for the effective charge corresponds
to the approximate solution of (\ref{betaf1}), but the omitted
terms lead only to the $\sim 1/N$ corrections to renormalon.
Again it is useful to expand $f(s)$ in the series
(\ref{serasym}). The corresponding recursion relation takes the
more complicated form
\begin{eqnarray}\label{recur1}
C_{k+1} &=& \fr{1}{k} \sum_{n=1}^k C_n C_{k+1-n}  \\
&+&\fr{\gamma}{k} \bigg[ C_k +2\sum C_n C_{k-n} +\sum C_n C_m
C_{k-n-m} \bigg] \nonumber \\
&-& \fr{\gamma^2}{k} \bigg[ \delta_{k,1} +3C_{k-1} + 3 \sum C_n
C_{k-n-1} +\sum C_n C_m C_{k-n-m-1} \bigg] \nonumber \ ,
\end{eqnarray}
where again $C_1=1$ and all $C$-s with nonpositive numbers are
supposed to be zero. After numerical simulation of
about $1000$ terms of the series (\ref{recur1}) we have found
\bq\label{numerics1}
R=0.13 \sum \left( \fr{9}{2} \alpha_0 \right)^N
N^{\fr{128}{81}} N!
 \ \ .
\ee

All the difference between the two definitions of the effective
charge (\ref{betaf}) and (\ref{betaf1}) lies in the different
choice of higher order corrections to the renormalization group
equation $b_2,b_3 \ ...$ . Thus the two asymptotics
(\ref{numerics}), (\ref{numerics1}) show explicitly that these
corrections make the contribution of the order of $\sim 1$ to
the overall normalization of renormalon, in complete agreement
with the conclusion of the preceding section.

{\bf Acknowledgements} Authors are thankful to A.~G.~Grozin,
M.~E.~Pospelov and A.~S.~Yelkhovsky for valuable discussions.
We would like also to thank V.~M.~Braun and A.~H.~Mueller for
useful comments on the preliminary version of the paper.
This work was supported by the Russian Foundation for
Fundamental Research under Grant 95-02-04607a.  The work of S.F.
has been supported by the INTAS Grant 93-2492 within the program
of ICFPM of support for young scientists.

\newpage
\begin{center}
{\bf Figure captions}
\end{center}

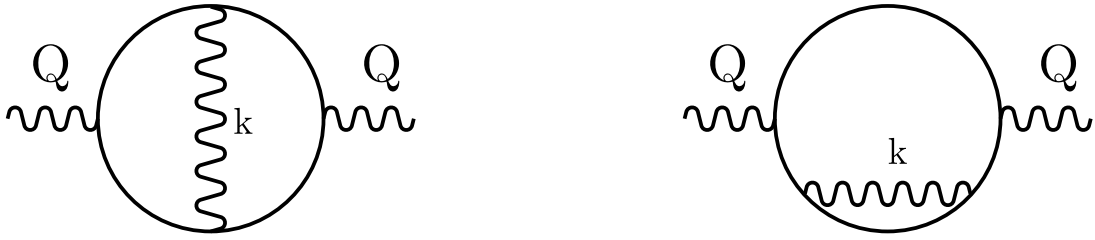
{}~~~~~{\bf Fig 1.} The renormalon-type graphs with exchange of
one soft gluon. The internal gluon line will be dressed in the
following figures.

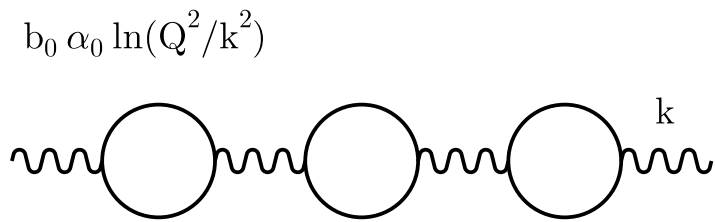
{}~~~~~{\bf Fig 2.} The example of the simplest chain of diagrams
corresponding to the renormalization of the soft gluon line.
Each bubble generates the factor $b_0 \alpha_0
\ln{\bigg(Q^2/k^2\bigg)}$.

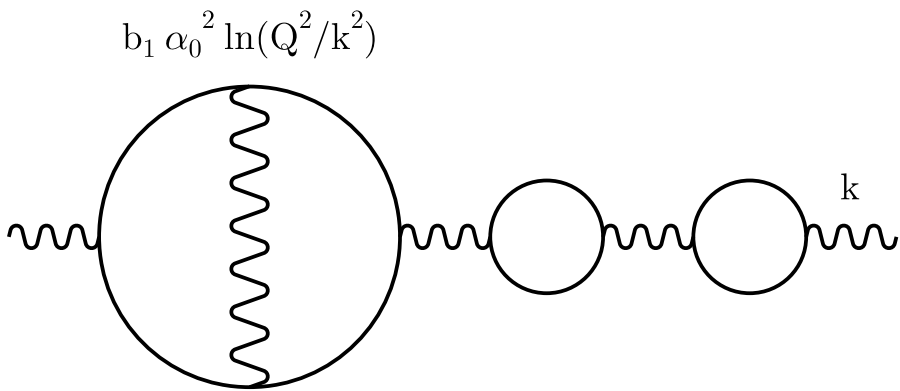
{}~~~~~{\bf Fig 3.} The example of the chain of diagrams with one
two -- loop insertion in soft gluon line. This second order
bubble generates the factor  $b_1 \alpha_0^2
\ln{\bigg(Q^2/k^2\bigg)}$.

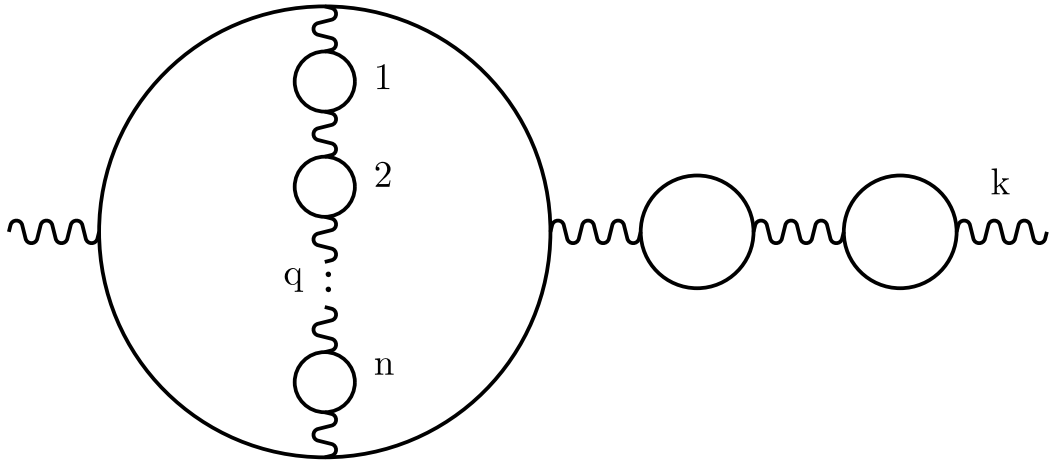
{}~~~~~{\bf Fig 4.} The dressing of the internal gluon line of the
second order bubble, shown in fig.~5, by $n$ simple bubbles.

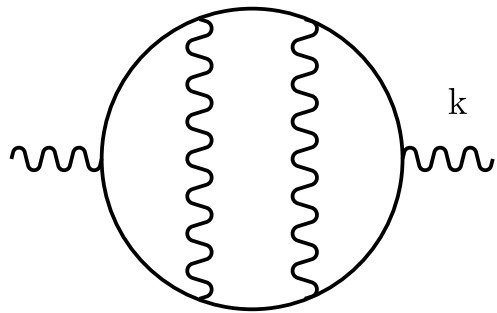
{}~~~~~{\bf Fig 5.} Three -- loop insertion to soft gluon line.
This diagram generate the factor  $b_2 \alpha_0^3
\ln{\bigg(Q^2/k^2\bigg)}$ and makes only $1/N$ correction
to the renormalon.

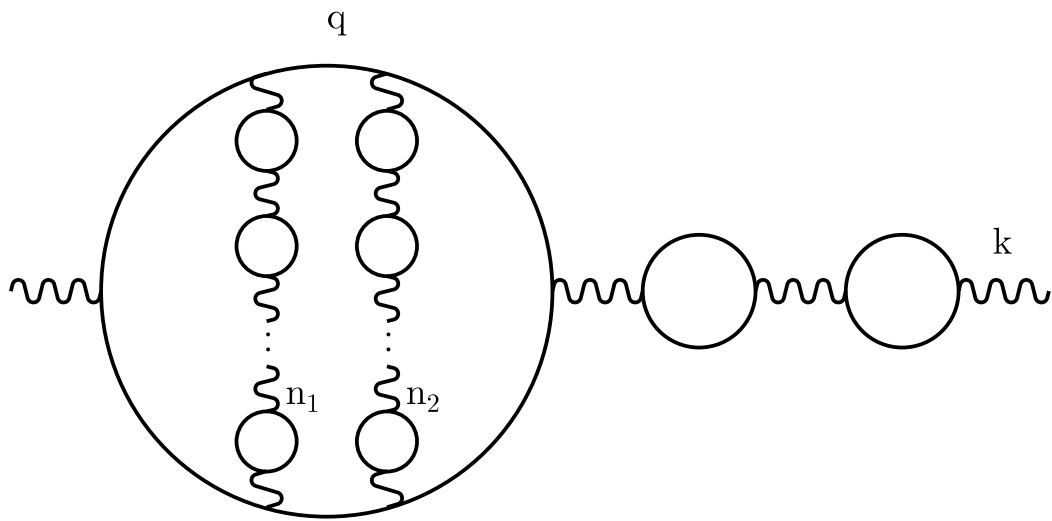
{}~~~~~{\bf Fig 6.} The same as in fig.~5, but with dressing of
two internal gluon lines by the simple chains of bubbles.  The
summation over $n_1$ and $n_2$ allows to compensate the extra
$\alpha$-s of the contribution of fig.~5, thus leading to the
correction of the order of $\sim 1$ to the renormalon.

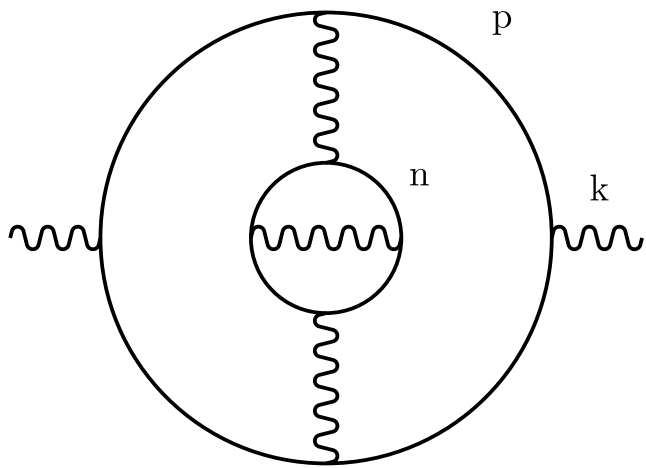
{}~~~~~{\bf Fig 7.} An example of the graph which corresponds to
the second iteration of two -- loop correction (see fig.~3).
Here $p$ is the number of (large) dressed bubbles which lie
along the soft gluon line and $n$ is the total number of
internal (small) dressed bubbles.

\end{document}